\documentclass[reprint,prb,superscriptaddress,showpacs]{revtex4-2}%

\usepackage{xr-hyper}
\setcitestyle{maxcitenames=6}
\usepackage{gensymb}
\usepackage{graphicx}
\usepackage{amsmath}
\usepackage{amsfonts}
\usepackage{amssymb}
\usepackage{framed}
\usepackage{float}
\usepackage[normalem]{ulem}
\usepackage{lipsum}
\usepackage[mathlines,switch]{lineno}
\providecommand{\U}[1]{\protect\rule{.1in}{.1in}}
\makeatletter
\renewcommand*{\fnum@figure}{{\normalfont\bfseries \figurename~\thefigure}}
\renewcommand*{\@caption@fignum@sep}{\textbf{ : }}
\makeatother
\usepackage{hyperref}
\hypersetup{
    colorlinks=true,
    linkcolor=blue,
    filecolor=magenta,      
    urlcolor=cyan,
}

\usepackage{siunitx}
\usepackage{adjustbox}
\usepackage{tabularx}
\usepackage[table]{xcolor}
\usepackage{booktabs}
\usepackage{gensymb}
\usepackage{xcite}

\makeatletter

\usepackage[normalem]{ulem}

\makeatother

\usepackage{hyperref}
\hypersetup{
    colorlinks=true,
    linkcolor=blue,
    filecolor=magenta,      
    urlcolor=cyan,
}
\begin{document}

\title{Near-degenerate Competing Magnetic Orders in EuAgAs: A Tunable Route to Altermagnetism}

\author{Mohamed El Gazzah}
\email{melgazza@nd.edu}
\affiliation{Department of Physics and Astronomy, University of Notre Dame, Notre Dame, IN 46556, USA}
\affiliation{Stavropoulos Center for Complex Quantum Matter, University of Notre Dame, Notre Dame, IN 46556, USA}
\affiliation{Center for Nanophase Materials Sciences, Oak Ridge National Laboratory, Oak Ridge, TN 37830, USA}

\author{Daniel Kaplan}
\affiliation{Center for Materials Theory, Department of Physics and Astronomy, Rutgers University, Piscataway, New Jersey 08854, USA}

\author{Zachary Morgan}
\affiliation{Neutron Scattering Division, Oak Ridge National Laboratory, Oak Ridge, Tennessee 37831, USA}

\author{Abhijeet Nayak}
\affiliation{Department of Physics and Astronomy, University of Notre Dame, Notre Dame, IN 46556, USA}
\affiliation{Stavropoulos Center for Complex Quantum Matter, University of Notre Dame, Notre Dame, IN 46556, USA}

\author{Resham Regmi}
\affiliation{Department of Physics and Astronomy, University of Notre Dame, Notre Dame, IN 46556, USA}
\affiliation{Stavropoulos Center for Complex Quantum Matter, University of Notre Dame, Notre Dame, IN 46556, USA}

\author{Sk Jamaluddin}
\affiliation{Department of Physics and Astronomy, University of Notre Dame, Notre Dame, IN 46556, USA}
\affiliation{Stavropoulos Center for Complex Quantum Matter, University of Notre Dame, Notre Dame, IN 46556, USA}

\author{Huibo Cao}
\affiliation{Neutron Scattering Division, Oak Ridge National Laboratory, Oak Ridge, Tennessee 37831, USA}

\author{Igor I. Mazin}
\affiliation{Department of Physics and Astronomy, George Mason University, Fairfax, VA 22030, USA}
\affiliation{Quantum Science and Engineering Center, George Mason University, Fairfax, VA 22030, USA}

\author{Nirmal J. Ghimire}
\email{nghimire@nd.edu}
\affiliation{Department of Physics and Astronomy, University of Notre Dame, Notre Dame, IN 46556, USA}
\affiliation{Stavropoulos Center for Complex Quantum Matter, University of Notre Dame, Notre Dame, IN 46556, USA}
\date{\today}

\begin{abstract}

Altermagnets (AMs) have recently emerged as a distinct magnetic class bridging central features of ferromagnets (FMs) and antiferromagnets (AFMs), offering new opportunities for spin-based electronics. While they possess zero net magnetization like collinear AFMs, they simultaneously exhibit momentum-dependent spin splitting long thought exclusive to FMs. Despite intense theoretical interest, experimentally accessible materials hosting both altermagnetism and nontrivial band topology remain scarce. EuAgAs, crystallizing in space group $P6_3/mmc$, was previously identified via density functional theory (DFT) as a bulk altermagnetic Dirac semimetal. Contrary to these predictions, our neutron diffraction experiments reveal that the bulk ground state adopts a $\mathbf{q} = (0,0,\tfrac{1}{2})$ AFM structure with an in-plane $\uparrow\uparrow\downarrow\downarrow$ spin sequence. Systematic DFT calculations, however, uncover a remarkable near-degeneracy among competing magnetic orders: the FM and AM configurations lie only $0.11$ and $0.40~\text{meV/f.u.}$ above the AFM ground state, respectively. We further show that while a simple Heisenberg model favors a spin-spiral ground state, the inclusion of non-Heisenberg biquadratic coupling stabilizes the observed commensurate AFM phase. This near-degeneracy renders the magnetic state highly tunable, with DFT predicting a transition to the altermagnetic phase under hydrostatic pressure at approximately $14 \text{ GPa}$, establishing EuAgAs as a controllable platform for accessing topological altermagnetism.

\end{abstract} 
\maketitle


\section{Introduction}

Altermagnets\cite{mazin_altermagnetism_2024} (AMs) have recently emerged as a distinct magnetic class that bridges central features of ferromagnets (FMs) and antiferromagnets (AFMs), reshaping the landscape of spin-based electronics\cite{smejkal_altermagnetic_2024,noauthor_altermagnets_nodate}. Like collinear AFMs, AMs possess zero net magnetization, suppressing stray fields and enabling ultrafast spin dynamics. At the same time, they exhibit electronic responses long thought to be exclusive to FMs, including a nonzero anomalous Hall effect\cite{jost_chiral_2025}, magneto-optical Kerr effect, and momentum-dependent spin splitting of electronic bands \cite{krempasky_altermagnetic_2024}. This unusual combination has positioned AMs as promising candidates for low-power, high-speed spintronic technologies \cite{noauthor_altermagnetism_nodate}.

The defining origin of AM lies in symmetry \cite{jungwirth_symmetry_2026,smejkal_altermagnetic_2024, Urru2025}. In conventional collinear AFMs, inversion or translational symmetries relate the two magnetic sublattices, enforcing Kramers degeneracy and spin-degenerate electronic bands. In contrast, AMs lack such mappings; instead, the sublattices are connected by rotational, mirror, or glide symmetries. This leads to spin-split electronic bands at generic momenta, with alternating spin polarization in momentum space—the hallmark of AM order  \cite{krempasky_altermagnetic_2024,noauthor_broken_nodate}.

\begin{figure*}
    \centering
    \includegraphics[width=.7\linewidth]{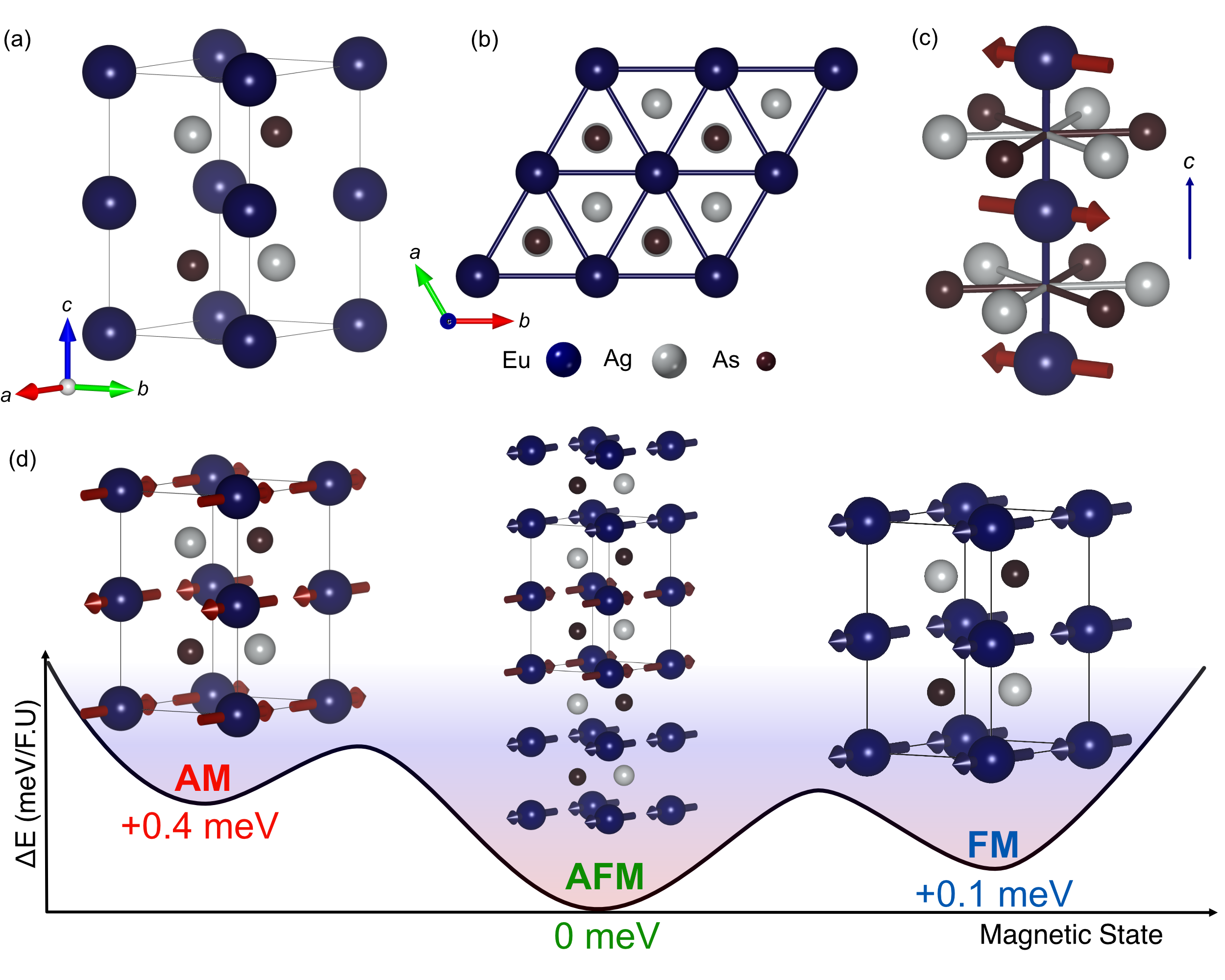}
    \caption{\textbf{Crystal structure and near-degenerate magnetic states of EuAgAs}
(a) Crystal structure of EuAgAs. (b) Top $c$-axis view of EuAgAs highlighting the hexagonal structure and the broken inversion symmetry between two Eu atoms along the $c$-axis (c) Altermagnetic  spins and the broken inversion symmetry between the antiferromagnetic Eu atoms. (d) Schematic energy landscape comparing several candidate magnetic configurations obtained from DFT. The antiferromagnetic (AFM) state represents the ground state, while the altermagnetic (AM) and ferromagnetic (FM) configurations lie slightly higher in energy, 0.4meV and 0.1meV respectively, indicating a near-degeneracy among competing magnetic states in EuAgAs.}
    \label{Fig1}
\end{figure*}

Beyond these functionalities, AMs have attracted rapidly growing attention in theoretical studies because of their natural compatibility with topological and relativistic band structures, such as Dirac and Weyl Fermions \cite{noauthor_altermagnetism_nodate,parshukov_topological_2025}. When combined, AM order and band topology are predicted to enable unconventional transport responses and symmetry-protected spin textures \cite{hu_topological_2024}. However, despite intense interest, the number of experimentally accessible materials that simultaneously host AM and nontrivial band topology remains extremely limited, significantly constraining progress toward materials design and device realization.

EuAgAs, which crystallizes in the BeZrSi-type structure with space group $P6_3/mmc$ (Fig. \ref{Fig1}(a)), provides a promising platform for realizing such a combination of phenomena. The highly symmetrical non-symmorphic space group $P6_3/mmc$ with the point group 6/mmm, makes EuAgAs an AM candidate  \cite{vsmejkal2022beyond}. The primitive cell contains two Eu atoms, making the putative AM state commensurate with the primitive cell. This is because the Eu atoms occupy the 2a Wyckoff positions (see Methods section, and Table \ref{ST1} and Fig. \ref{FigS1} in the Supplementary information (SI)), which are related by $6_3$ screw symmetry in the space group $P6_3/mmc$.

From a more intuitive perspective, the structure can be viewed as consisting of three Eu layers within a unit cell, where the bond center between Eu atoms in adjacent layers lacks inversion symmetry, as illustrated in Fig. \ref{Fig1}(c). A collinear antiferromagnetic ordering with $\mathbf{q}$ = (0,0,0), as shown in Figs. \ref{Fig1}(c) and (d), renders this compound AM by definition. Prior density functional theory (DFT) calculations have identified this magnetic configuration as the ground state \cite{Jin2021,Laha2021}. Those studies further indicated that, in this state, the material hosts Dirac semimetallic behavior. It is important to note that AM spin splitting in materials with space group $P6_3/mmc$, including MnTe\cite{noauthor_broken_nodate,amin_nanoscale_2024}, CrSb\cite{willis_crystal_1953,zeng_observation_2024,zhou_manipulation_2025}, and CoNb$_4$Se$_8$\cite{noauthor_altermagnetism_nodate-1,dale_relativistic_2026}, occurs at finite $k_z$. Consequently, earlier calculations that considered only the $k_z$ = 0 direction overlooked this AM splitting. In the ferromagnetic state, accessible with an applied magnetic field at approximately 3 T, the Dirac point splits into a pair of Weyl points. 

Taken together, these results position EuAgAs as a compelling platform where AM, Dirac, and emergent Weyl physics may intertwine. However, establishing and interpreting these phenomena critically depends on a precise determination of the underlying magnetic structure, since the symmetry of the ordered state dictates whether AM behavior can arise.

In this study, we have carried out neutron diffraction experiments to determine the magnetic ground state of EuAgAs, revealing a $\mathbf{q}=(0,0,1/2)$ antiferromagnetic structure (Fig. 1(d), middle panel). While this magnetic configuration does not support bulk AM, it leaves open the intriguing possibility of symmetry-lowered surface AM. 
More intriguingly, density functional theory (DFT) calculations show that the energy differences among the experimentally observed antiferromagnetic (AFM) state, the AM configuration, and the ferromagnetic (FM) state are extremely small (less than 0.5 meV) [Fig. \ref{Fig1}(d)], placing the system near a magnetic instability. This near-degeneracy renders the magnetic ground state highly tunable, with even modest perturbations sufficient to drive transitions between these competing phases. These results establish EuAgAs as a unique material in which multiple magnetic phases hosting topological electronic band features can be accessed and controlled through external tuning parameters such as pressure or chemical doping.


\begin{figure*}
    \centering
    \includegraphics[width=.7\linewidth]{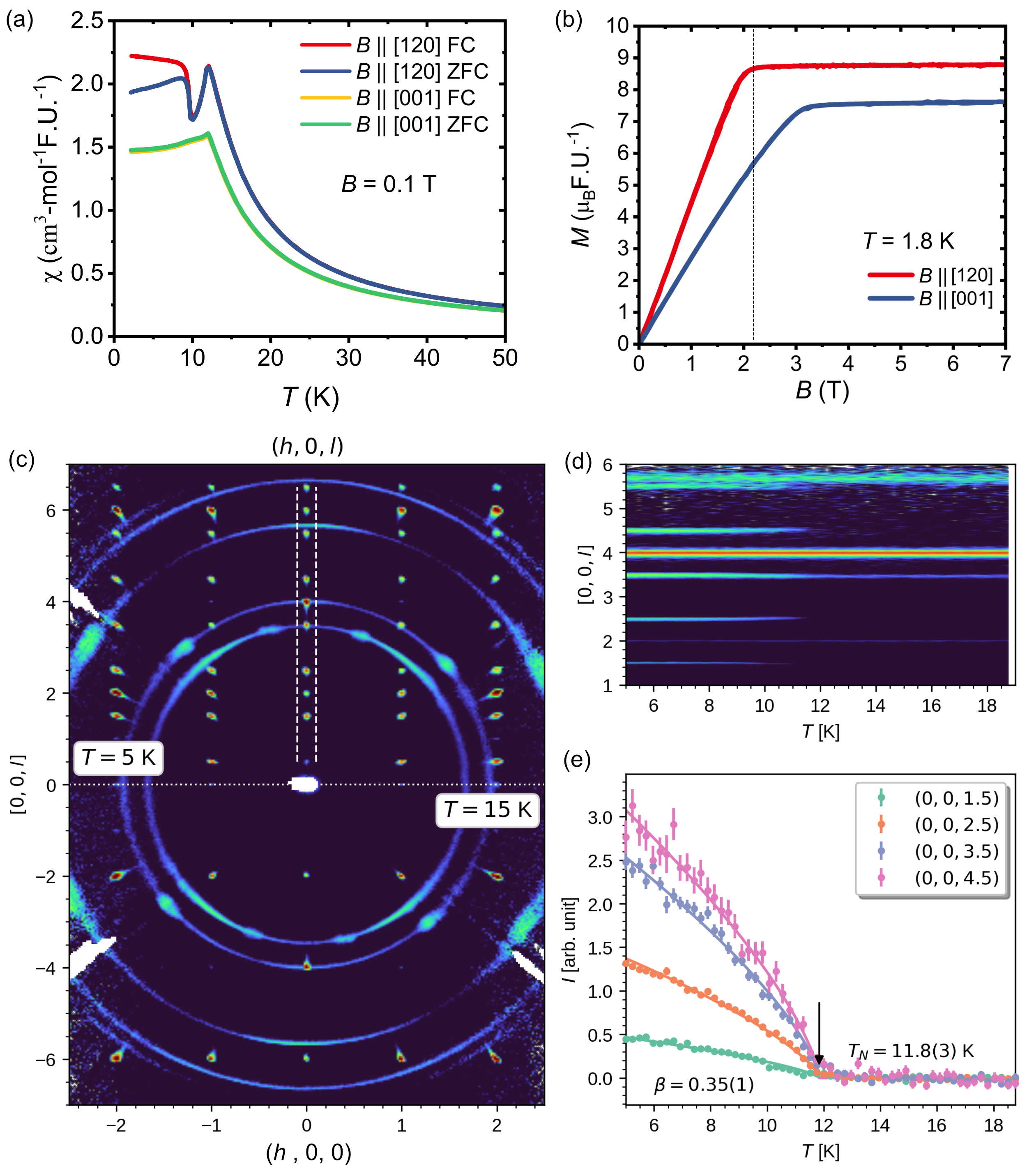}
    \caption{\textbf{Magnetic and neutron-scattering signatures of antiferromagnetic order in EuAgAs.}
    (a) Temperature dependence of the magnetic susceptibility $\chi$(T) measured under zero-field-cooled (ZFC) and field-cooled (FC) conditions for an applied field of $0.1$ T along the $[001]$ and $[120]$ crystallographic directions. (b) Field-dependent magnetization $M(B)$ measured at $T = 1.8$ K for magnetic fields applied along $[001]$ and $[120]$. (c) Reciprocal-space maps of the $(h,0,l)$ plane measured above the magnetic transition ($T=15$ K, lower hemisphere) and at base temperature ($T=5$ K, upper hemisphere). (d) Temperature evolution of the scattering intensity along the $[0,0,l]$ direction. (e) Temperature dependence of the magnetic peak intensities at $l = 1.5, 2.5, 3.5,$ and $4.5$, indicating an ordering temperature of $T_\mathrm{N} \approx 11.8$ K.}
    \label{Fig2}
\end{figure*}

\section{Results and Discussion}

EuAgAs orders antiferromagnetically below 12 K, as evidenced by the susceptibility data (Fig. \ref{Fig2}(a), and more details in Fig. \ref{FigS2} of SI). A forced ferromagnetic state is achieved with a modest magnetic field ($B$) of 3 T at 1.8 K, as shown in Fig. \ref{Fig2} (b), which also shows an easy-plane magnetic anisotropy (further MH curves are presented in Fig. \ref{FigS3} of SI). Figs. \ref{Fig2}(c-e) present neutron diffraction measurements of EuAgAs, revealing the development of long-range magnetic order below $T_\mathrm{N}$. Measurements were performed at the CORELLI beamline at the Spallation Neutron Source (SNS), Oak Ridge National Laboratory (see Methods Section for details). The $(h,0,l)$ plane measured at $T=15$~K (lower half of Fig.~\ref{Fig2}(c)) shows only nuclear Bragg reflections, consistent with the crystallographic structure. Upon cooling to $T=5$~K (upper half of Fig.~\ref{Fig2}(c)), additional satellite peaks emerge at half-integer $l$-positions, i.e., $(0,0,l+\tfrac{1}{2})$. These reflections are absent above the transition and therefore arise from magnetic ordering. Their periodicity directly identifies a propagation vector $\mathbf{q} = (0,0,\tfrac{1}{2})$, corresponding to a doubling of the magnetic unit cell along the $c$-axis, in contrast to prior DFT predictions of $\mathbf{q} = (0,0,1)$, and indicating antiferromagnetic stacking of Eu moments between adjacent layers.

A cut along the $[0,0,l]$ direction (Fig.~\ref{Fig2}(d)) isolates the magnetic scattering, clearly separating nuclear peaks at integer $l$ from magnetic peaks at half-integer positions. The appearance of well-defined magnetic Bragg peaks below $T_\mathrm{N}$ confirms the formation of long-range magnetic order.

The temperature evolution of the magnetic scattering (Fig.~\ref{Fig2}(d, e)) shows a continuous onset  of intensity below $T_\mathrm{N} \approx 12$~K. The integrated intensities of the magnetic reflections at $l = 1.5, 2.5, 3.5,$ and $4.5$ follow a consistent order-parameter-like behavior. A power-law fit yields a critical exponent $\beta \approx 0.35$, consistent with a second-order phase transition in a three-dimensional magnetic system. The identical temperature evolution across multiple reflections confirms that they originate from a single magnetic phase and establishes the bulk nature of the transition.

The presence of magnetic intensity at $(0,0,l+\tfrac{1}{2})$ positions places a strong constraint on the moment orientation. Since neutron magnetic scattering is sensitive only to components of the magnetic moment perpendicular to the scattering vector $\mathbf{Q}$, reflections along $(0,0,l)$ probe moments lying within the $ab$-plane. The observation of strong magnetic intensity at these positions therefore rules out moments aligned along the $c$-axis.

Refinement of candidate magnetic models, guided by symmetry analysis, further constrains the moment direction within the plane. The best agreement with the observed intensities is obtained for a model in which the Eu moments align along the crystallographic $a$-axis. Combined with the propagation vector $\mathbf{q} = (0,0,\tfrac{1}{2})$, this yields a magnetic structure consisting of ferromagnetically aligned Eu moments within each layer, stacked along the $c$-axis in an antiferromagnetic $\uparrow\uparrow\downarrow\downarrow$ sequence. The refined ordered moment is consistent with the expected value for Eu$^{2+}$.

The neutron diffraction results thus establish a bulk magnetic ground state that departs markedly from prior theoretical predictions, motivating a reexamination of the microscopic magnetic interactions in EuAgAs. To reconcile the experimentally observed magnetic structure and the prior DFT predicted magnetic ground state, we performed systematic DFT calculations with and without spin-orbit coupling (SOC). All ground state energy calculations are carried out with VASP \cite{kresse1996efficient,kresse1996efficiency} using PAW pseudopotentials \cite{kresse1999ultrasoft}. In order to validate that the near-degeneracy is not an artifact of the pseudopotential method, we have also carried out limited AE (all-electron) calculation on our principal candidates using WIEN2k \cite{blaha2020wien2k}. These calculations not only clarify the origin of the observed ground state but also reveal the remarkable near-degeneracy among the AFM, AM, and FM phases that defines the central result of this work.

\begin{figure*}
    \centering
    \includegraphics[width=1\linewidth]{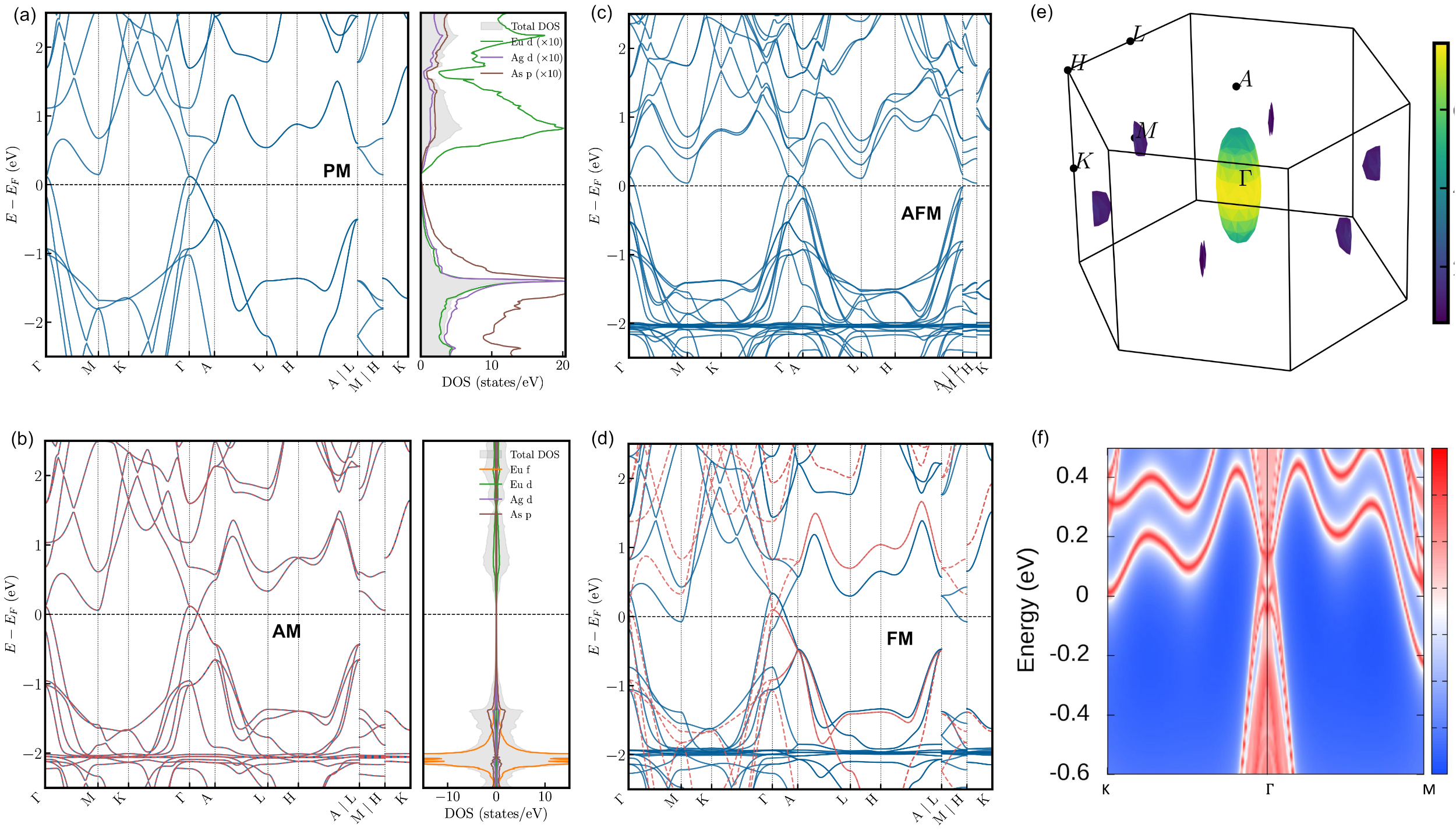}
    \caption{\textbf{Electronic structure and topological features across magnetic phases of EuAgAs.}
    (a) Bands of the paramagnetic state, without SOC, in the space group P6$_3$/mmc, with Eu-f bands frozen in core. Note the band crossing along $\Gamma-A$. The density of state in the conduction band is dominated by Eu-d and Ag-d orbitals. The valence states near $E_F$ are predominantly of As-p character. (b) Band structure, without SOC of the altermagnetic state (A-type AFM). Dashed and solid lines indicate spin $\uparrow$ and $\downarrow$, respectively. The lack of spin splitting is due the non-symmorphic symmetry-enforced protection of the $k_z =0, 1/2$ planes, which also form the high-symmetry lines in this space group (described in the main text). (c) Band structure, with SOC, of the AFM ((0,0,1/2) state). All bands are doubly degenerate due to the $\mathfrak{t}_{1/2}T$ symmetry of this structure. The crossing near $\Gamma-A$ [compare it with panel (a)] is lifted, with a gap of $1.05$ meV. (d) Band structure of the FM state showing the lifting of spin degeneracy. The previously quadruply degenerate crossing along $\Gamma-A$ is split into two Weyl points. (e) Three-dimensional Fermi surface of FM EuAgAs showing a large elliptical pocket at $\Gamma$ and smaller, fragmented pockets near the $M$ points. These contribute anisotropic SdH oscillations. (f) Surface states on the (001) surface in the AFM phase, indicating the emergence of topological states in the vicinity of the (now gapped, with SOC) crossing along $\Gamma-A$.}
    \label{fig:bands}
\end{figure*}

\begin{figure*}
    \centering
    \includegraphics[width=.7\linewidth]{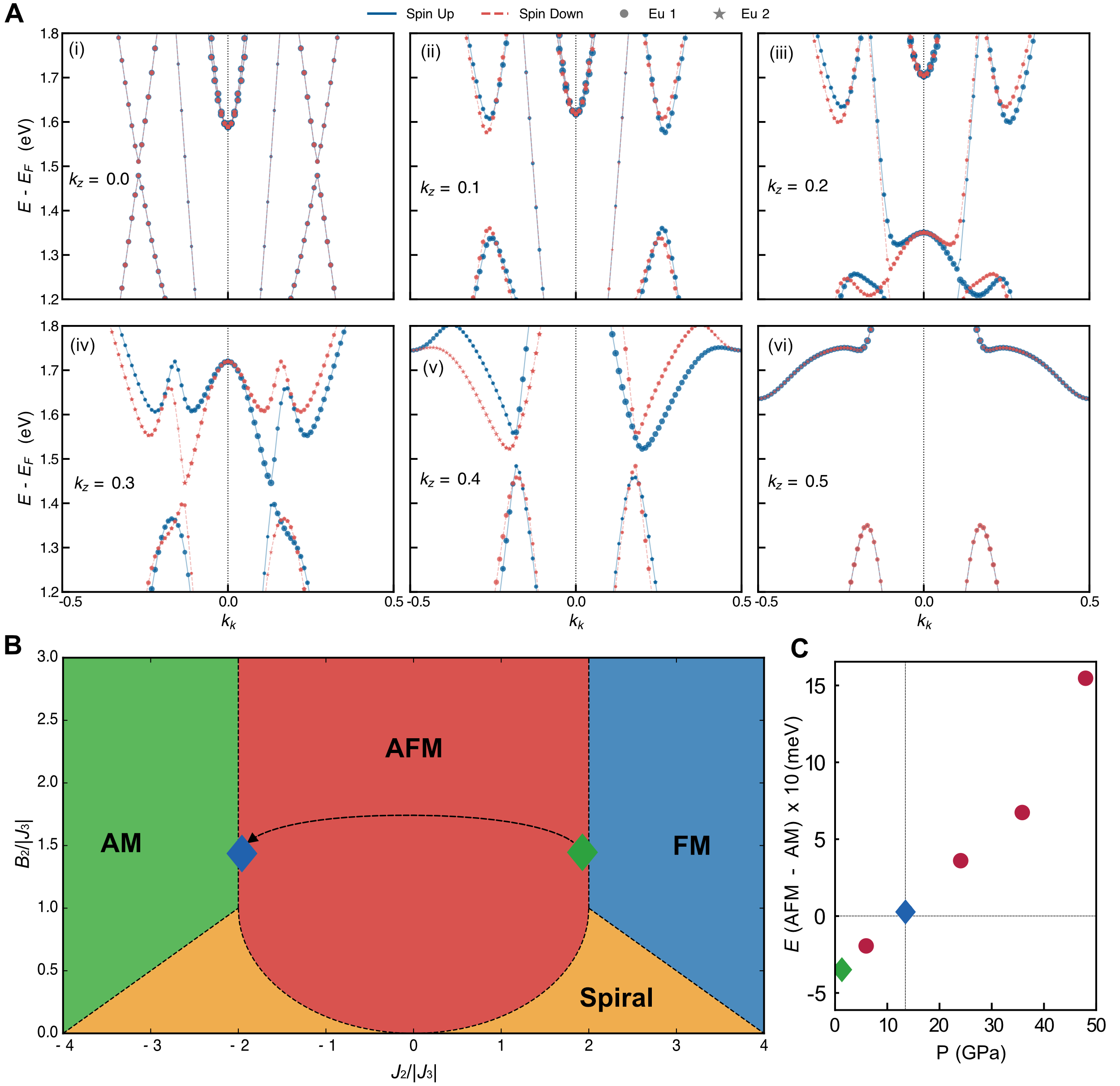}
    \caption{\textbf{Altermagnetic spin splitting, magnetic phase diagram and pressure tunability in EuAgAs.} (A) The band structure in the AM phase, with the specific projection on both \textit{spin} (blue solid, and orange dashed lines, respectively) and Eu sublattice (circles and stars). Bands are plotted along the cut $k_y =0, k_x$ with finite $k_z$, where (see symmetry discussion) spin splitting is expected. The energy window here is taken away from $E_F$ in order to capture the Eu-d bands which are strongly coupled to magnetism. As expected by our symmetry analysis, the bands are spin split only for finite $k_z$, whose evolution from $0$ to 1/2, is shown in panels (i)-(vi). (b) Magnetic phase diagram as a function of the Heisenberg parameters $J_2,J_3$ and the biquadratic exchange $B_2$. Green and blue diamonds denote the equilibrium and finite pressure (14 GPa) parameters, respectively.     
    (c) Evolution of the ground state energy relative to the AFM, as a function of pressure in DFT. We find a degeneracy between AM and AFM at roughly $14$ GPa.}
    \label{fig:2}
\end{figure*}

We considered several simple collinear magnetic configurations within the primitive cell, including the experimentally observed AFM structure with $\mathbf{q}=(0,0,1/2)$, the A-type AFM configuration that realizes the AM state, and FM states with moments oriented along both the $a$- and $c$-axes (see Fig. \ref{fig:2} for the schematic of configurations). We begin by analyzing the paramagnetic electronic structure, treating the Eu-$4f$ electrons as frozen core states [Fig.~\ref{fig:bands}(a)]. The density of states near the Fermi energy is dominated primarily by Eu-$d$ and Ag-$d$ orbitals. Owing to the large localized moment associated with the Eu$^{2+}$ ion, SOC is expected to produce only weak magnetic anisotropy. Indeed, our calculations reveal only a very small magnetocrystalline anisotropy energy ($\sim 0.07$ meV/f.u. for the (0,0,1/2) AFM), favoring moment alignment along the $a$-axis, consistent with the neutron diffraction results and prior DFT work~\cite{Jin2021}. The plots in Fig.~\ref{fig:bands} do not include SOC with the exception of Figs.~\ref{fig:bands}(c, f), where SOC was used to resolve small gaps related to topology.

The calculated energies of the various magnetic configurations are summarized in Tab~\ref{tab:1}, relative to the experimentally observed AFM ground state. Strikingly, the FM state with moments along the $a$-axis lies only $\sim 0.11$ meV/f.u. above the ground state, while the AM configuration is higher by only $\sim 0.4$ meV/f.u. The extremely small energy differences among the experimentally observed AFM, FM, and AM states place EuAgAs in a highly frustrated, meta-magnetic landscape, indicating that modest perturbations could readily tune the system between these competing phases. This near degeneracy establishes EuAgAs as a promising platform for engineering emergent magnetic and topological states, including AM, through external tuning parameters such as hydrostatic or chemical pressure. Motivated by this possibility, we further performed pressure-dependent calculations to investigate the evolution and stability of these nearly degenerate magnetic phases. We first discuss the corresponding electronic structures of each magnetic configuration and subsequently examine the energetic evolution under pressure.

\begin{table}[htbp]
    \centering
    \caption{Magnetic energy (with SOC) of  in \textbf{meV/f.u.} from first-principles}
    \label{tab:1}
    \begin{tabular}{@{} c c c @{}} 
        \toprule
        \textbf{Magnetic state } & VASP & AE \\
        \midrule
        AM & 0.402  & 0.8 \\
        FM-x & 0.11 & 0.1  \\
        AFM $(0,0,1/2)$ & 0.0 & 0.0\\ 
        FM-z & 1.05 & \\
        \bottomrule
    \end{tabular}
\end{table}

The band structure of the experimental AFM magnetic structure is plotted in Fig.~\ref{fig:bands}(c), and as stated above, the bands in Fig.~\ref{fig:bands}(c) are computed with SOC. Due to the symmetries of the extended AM in the (0,0,1/2) cell, the combined action of half-translation and time-reversal renders all bands $\mathfrak{t}_{1/2}T$ locally Kramers degenerate and indeed the band structure is found to be doubly-degenerate everywhere in the Brillouin zone, as we elaborate on, below. 

Since the application of perpendicular magnetic field leads to complete spin polarization at a modest field of 3 T [see Fig. \ref{Fig2}(b)], we also consider the primitive FM configuration (without SOC) and its band structure [Fig.~\ref{fig:bands}(d)] which reveals the lifting of the four-fold degenerate crossing along $\Gamma-A$ into two separate Weyl points. One of the crossings is related to the intersection of majority/minority spins and gaps out upon introduction of SOC (see Fig. \ref{FigS4}).

From this structure, we find complex fermiology that varies strongly with chemical potential. Near the calculated $E_\text{F}$, we present the full, 3D Fermi surface in Fig.~\ref{fig:bands}(e). Both the large and elliptical $\Gamma$ pocket and the electron pockets near $M$ shown in Fig.~\ref{fig:bands}(e) will likely vary strongly with tilting the magnetic field and contribute to anisotropic quantum oscillations, which can be readily accessed by analyzing Shubnikov–de Haas (SdH) and/or de Haas–van Alphen (dHvA) oscillations which we observe appearing at 8 T (see Fig. \ref{FigS5}). However to fully map out this complex Fermi surface, it may require high magnetic field measurements.

Finally, due to its close energetic proximity to the experimentally realized ground state, we discuss the AM phase, which may be accessible through suitable physical or chemical tuning. The band structure and DOS for this phase is plotted in Fig.~\ref{fig:bands}(b). 
We now discuss the nature of the non-relativistic spin splitting in the AM phase. The spin splitting in the A-type AFM phase (AM phase) is supported by the fact that the inversion center in the primitive cell coincides with a Eu $2a$ Wyckoff position. Without SOC, different Eu sublattices are related via $\mathbf{R}|C_{2z}$, where $\mathbf{R}$ is a lattice operation and $C_n$ acts on the spin sector only \cite{vsmejkal2022beyond}. 

As a result, non-relativistic spin-splitting is permitted, with the following nodal planes enforced by the non-symmorphic symmetry. Within P6$_3$/mmc, consider $\mathfrak{g} : (x,y,z)\to(y,x,z+1/2)$. This symmetry connects the two Eu $2a$ sublattices; it therefore also acts as a ``spin flip" operation. For any $k$ point in reciprocal space $(k_1,k_2,k_3)$, $\mathfrak{g} \cdot\mathbf{k} = (k_2,k_1,k_3)$. Now, if we consider the spin-splitting function $\Delta(\mathbf{k}) = E_{\uparrow}(\mathbf{k}) -E_{\downarrow}(\mathbf{k})$, $\mathfrak{g}\cdot \Delta = -\Delta(\mathfrak{g} \cdot \mathbf{k})$. For this operation, the line $(k_1,k_1, k_3)$ (e.g. $\Gamma -K$ on the $k_z =0$ plane) is enforced to have $\Delta = 0$. Now, in similar fashion, consider $(x,y,z) \to (-x,-x+y,z+1/2), (x-y,-y,z+1/2)$. These are the remaining spin symmetries, and along high symmetry directions, map $(u,0,k_z) \to (-u,0,k_z)$ (for the former) and $(u,-u,k_z)$ for the latter. This means that the $\Gamma-M$ line along any cut where $k_z \neq 0,1/2$ is not enforced to be zero. 

Indeed, we plot precisely this segment along $k_x, k_y=0$ for several values of $k_z$ in Fig.~\ref{fig:2}(a), without SOC. We show a slice of the band structure away from $E_F$, as the states near $E_F$ are predominantly Ag-$d$, As-$p$ and these do not hybridize strongly enough (without SOC) with Eu orbitals. As explained, we observe the emergence of non-relativistic spin splitting along $k_x$ whenever $k_z \neq 0,1/2$, demonstrating the symmetry argument we outlined earlier. 

Here, the spin-splitting is \textit{odd} in the plane due to the non-symmorphic $C_{2z}$, $(x,y,z)\to(-x,-y,z+1/2)$. In fact, we are able to distinctly show this by examining the projection of the bands on the two Eu sublattices, in Fig.~\ref{fig:2}. The dominant contribution flips when $k_x \to -k_x$, that is from Eu-1 to Eu-2, as expected. For particular values of $k_z$, we find a spin splitting as large as $100~\textrm{meV}$. This shell is predominantly composed of Eu-d orbitals. Since the hybridization here is non-relativistic, it relies on coupling between $f$ and $d$ electrons via a Kondo coupling \cite{Zhao2025}. With the addition of $z \to -z$ non-symmorphic operations, the symmetry of this AM state is indeed $g$-wave, as stated in Ref.~\cite{vsmejkal2022beyond}. Consequently, this marks one of the few examples of a Kondo lattice-driven g-wave AM.

In the AFM (0,0,1/2) state, by contrast, the space group is Amm2.$1'_a$ which maintains a $1'$ operation. This is a $\mathfrak{t}_{1/2} T$ where $\mathfrak{t}_{1/2}: (x,y,z)\to(x+1/2,y+1/2,z+1/2)$ and thus enforces Kramers degeneracy at every $k$-point. Incidentally, we do not observe any significant forces when introducing SOC to the calculation. As a result, we predict the orthorhombic distortion in this case to be negligible. 

Finally, the ferromagnet along $z$, achieved with out-of-plane magnetic field, would retain most of the symmetries of the underlying lattice. The magnetic point group is 6/mm$'$m$'$ and preserves the key $6_3$ screw symmetry. As shown in Fig.~\ref{fig:bands}(D), the effect of magnetic order (without SOC) is to split the Dirac point into doubled Weyl points, protected by the non-symmorphic symmetry. These states are themselves topological \cite{Jin2021}, and as this phase is stabilized even with moderate magnetic fields, an accurate experimental investigation of its properties is left to future work. 

To accurately capture the competition between the magnetic phases we considered in Tab.~\ref{tab:1}, we map the spin state using a Heisenberg model involving nearest neighbor interactions. To a good approximation, the intralayer coupling $J_1$ is ferromagnetic and for all A-type antiferromagnets does not affect the out-of-plane interaction. This renders our model effectively 1D along $z$. To leading order,
\begin{align}
    H = H_{FM}-J_2\sum_{\langle i,j\rangle}\mathbf{S}_i\cdot\mathbf{S}_j-J_3\sum_{\langle\langle i,j\rangle\rangle}\mathbf{S}_i\cdot\mathbf{S}_j.
    \label{eq:heis}
\end{align}
Where $\langle ...\rangle$ and $\langle\langle...\rangle\rangle$ stand for nearest and next-nearest neighbor interactions, respectively. For the Eu$^{+2}$ state, we take $S=7/2$. After Fourier transforming, we find the the magnetic interaction to be $-J(q_z)/2 = J_2\cos(q_zc/2)+J_3\cos(q_zc)$ (for a doubled unit cell). Maximizing this (note our sign convention in Eq.~\eqref{eq:heis}) with respect to $q_z$, we find that $q_z = (2/c)\arccos(-J_2/(4J_3))$, clearly favoring a spin-spiral state, unless $J_2 = 0$ or $|J_2|/(4|J_3|) > 1$. For the latter case and $J_2<0$, $q_z =0$ is preferred, and would yield the A-type AFM, generically. To explain, however, the emergence of the doubled-unit cell state, a common mechanism that stabilizes collinear, commensurate order is biquadratic interactions \cite{kartsev2020biquadratic,slonczewski1991fluctuation,fedorova2015biquadratic}. This is especially true for magnets with weak exchange, as we argue is the case here. The biquadratic interactions take the form,
\begin{align}
    H' = H-B_2\sum_{\langle i,j\rangle}(\mathbf{S}_i \cdot \mathbf{S}_j)^2.
\end{align}
Such a term strongly penalizes non-collinear orderings, which give rise to spirals. We test the effect of this term by proposing a generic spiral ansatz $\mathbf{S}_n/S = \hat{x}\cos(nqc/2)+\hat{y}\sin(nqc/2)$. $n$ here denotes the $z$ plane of the Eu sublattice. It is clear that $q = 0$ and $q=2\pi/c$ are the FM and AM, respectively. In this model energy of the (0,0,1/2) AFM is fixed at $E=J_3-B_2S^2$. It also follows that (with standard identities) $\mathbf{S}_n \cdot\mathbf{S}_{n+1} = \cos(qc/2)$ and similarly for higher order neighbors. The energy equation for the spiral becomes, $E/S^2 = J_3-J_2\cos(\theta)-(2J_3+B_2S^2)\cos^2(\theta)$, and $\theta=qc/2$.
We now plot the full phase diagram of this system, and for simplicity absorb $J_i \to J_i/S^2$ and $B_2 \to B_2/S^4$, as we show in Fig.~\ref{fig:2}(b). This phase diagram captures all nearly-degenerate orders found in first principle studies: FM, AFM, the putative spiral order and the AM, provided that $J_3 < 0$. Then, the phase diagram is split into two regions, controlled by the sign of $J_2$. For positive $J_2$ (note the sign convention in Eq.~\eqref{eq:heis}), the system borders the FM phase. If $J_2$ is negative the system favors the commensurate AM (i.e., $q_z=0$), provided $B_2\neq 0$. From first-principles, we directly extract the values of $J_2,J_3,B_2$ from the energy difference between associated magnetic structures. We find $J_2 = 0.1\textrm{meV}, J_3 = -J_2/1.95$ and $B_2 = 1.52 J_3$ after appropriately normalizing by $S^2$. We use $J_2$ to estimate $T_N \approx S(S+1)J_2/(3k_B) = 6.09 $K, comparable to the experimental $T\approx 12K$. The 1D character of our model realizes an intriguing possibility related to the dependence of the vertical separation between Eu sites. Since the intralayer coupling will not be significantly affected by out-of-plane contraction, we apply pressure to our unit cell. From DFT, we find a transition from the AFM to the AM at roughly 14GPa, as we show in Fig.~\ref{fig:2}(c). Correspondingly, near the transition, we observe a shift in the Heisenberg interaction which translates into a sign change of $J_2$. This is again consistent with the notion that magnetism here is RKKY-dominated \cite{elliott2013magnetic,ruderman1954indirect,kasuya1956theory,yosida1957magnetic}, and thus strongly depends on the precise structure of the Fermi surface, which can be affected by lattice spacing, even for moderate pressures. In Fig.~\ref{fig:2}(b) this transition is illustrated by a shift \textit{across} the phase diagram, namely, from the green diamond (equilibrium) to the blue diamond ($\approx 14$ GPa), changing the overall instability. Accessing the spiral states through a similar approach is left for future work.

\section{Conclusion}
In summary, neutron diffraction establishes that the bulk ground state of EuAgAs is a \textbf{q} = (0, 0, 1/2) AFM arrangement, with an in-plane $\uparrow\uparrow\downarrow\downarrow$ spin arrangement, rather than the AM phase anticipated from prior theoretical work. Nevertheless, the DFT energy landscape reveals a near-degeneracy in magnetic states: the AM and FM configurations lie only 0.40 meV and 0.11 meV per formula unit above the ground state, respectively. Notably, the Heisenberg model alone favors a spin-spiral ground state over the $\uparrow\uparrow\downarrow\downarrow$ order; it is the inclusion of non-Heisenberg biquadratic coupling that stabilizes the observed commensurate AFM phase. This near-degeneracy is key, as it transforms EuAgAs from a failed AM candidate into a uniquely tunable one. The proximity of nearby magnetic orders is also a hallmark of other rare-earth magnets, especially given the presence of topological states \cite{kushnirenko2022rare,huang2024hidden,schrunk2022emergence}.
Chemical doping, dimensional confinement in thin films with an even number of layers, and hydrostatic pressure each provide realistic experimental routes for crossing into the AM phase (and other phases), with DFT predicting an AFM-to-AM transition near 14 GPa. Furthermore, given the sensitivity of the energetics within EuAgAs, the AM phase can potentially be achieved through a simple applied magnetic field. This places EuAgAs uniquely at the confluence of topology, AM and spiral magnetic phases.

\section{Materials and Methods}

\subsection{Single Crystal Growth}

Single crystals of EuAgAs were grown by the flux-method using Bi as flux. Eu pieces, Ag pieces, As chunks and Bi shot were added into a 2 mL aluminum oxide crucible in molar ratio of 1:1:1:10. The crucible was then heated in a fused silica ampule under vacuum. The sealed ampule was heated to 1100 $^{\circ}$C and kept at this temperature for 24 hours, and then cooled to 700 $^{\circ}$C at a rate of 1.5 $^{\circ}$C/hour. Once the furnace reached 700 $^{\circ}$C, the tube was decanted using a centrifuge to separate the Bi flux keeping the EuAgAs plate like crystals.

\subsection{Single Crystal X-Ray Diffraction}

Structural characterization was performed using single-crystal X-ray diffraction (XRD) at room temperature. Data were collected on a well-faceted crystal, using a Bruker Quest diffractometer equipped with a Bruker PHOTON III detector, and employing a combination of $\omega$- and $\phi$-scans with a step size of 0.5 degree. The diffraction data were corrected for absorption and polarization effects and analyzed to determine the appropriate space group. Structure solution was carried out using dual-space algorithms, followed by routine expansion of the initial model. Final structural refinement was performed using full-matrix least-squares analysis against $F^2$ values for all observed reflections \cite{sheldrick_shelxt_2015}. The extinction coefficient was also refined.

\subsection{Magnetic Measurements}

Magnetic properties were measured using a 7-T Quantum Design Magnetic Property Measurement System (MPMS3) over a temperature range of 1.8 to 400 K using a VSM mode with a peak amplitude of 8 mm and averaging time of 2 seconds. A plate-like single crystal was mounted on a quartz holder using GE varnish for in-plane measurements, while a brass holder was used to measure the out-of-plane direction. Magnetization and magnetic susceptibility were measured along both the [120] and [001] axes after confirming the crystal orientation through Laue diffraction.

\subsection{Electronic Transport Measurements}

DC magnetization, resistivity, and heat capacity measurements were carried out using two separate Quantum Design Dynacool Physical Property Measurement Systems (PPMS) equipped with 9 T and 14 T magnets. The ACMS II option was used for DC magnetization measurements. Single crystals of EuAgAs were polished to appropriate dimensions for electrical transport measurements. Crystal orientation along the [100], [010], and [001] directions was determined using a Photonic Science X-ray Laue diffractometer, along with the Cologne Laue Indexation Program to refine and resolve the crystallographic axes. Resistivity and Hall effect measurements were carried out using the standard four-probe method. Platinum wires (25 $\mu$m in diameter) were used for electrical contacts, with contact resistances below 20 $\Omega$. The electrical contacts were affixed using Epotek H20E silver epoxy, and a 2.5 mA current was applied during transport measurements. To correct for contact misalignment, magnetoresistance and Hall resistivity data were symmetrized and anti-symmetrized, respectively. MR is defined as $MR = (\rho(H)-\rho(0))/\rho(0)$, where $\rho(H)$ and $\rho(0)$ are the longitudinal resistivity with and without an applied magnetic field, H. The Hall resistivity was asymetrized from the positive and negative applied magnetic fields via $\rho_{\text{H}}=[\rho_\text{T}(+H)-\rho_\text{T}(-H)]/2$, where $\rho_\text{T}$ is the resistivity measured via the transverse voltage contacts in the Hall bar geometry.

\subsection{Neutron Diffraction}

Neutron diffraction measurements were performed at the CORELLI beamline at the Spallation Neutron Source (SNS), Oak Ridge National Laboratory\cite{ye_implementation_2018}. A plate-like single crystal (3 × 2 × 1 mm³) was mounted on an aluminum pin using cyanoacrylate adhesive and installed in a closed-cycle refrigerator. The sample was cooled to a base temperature of 4.4 K.

Reciprocal-space volumes were collected by rotating the crystal about the single-axis goniometer in 6° increments with a counting time of 6 min per angle. The resulting datasets were combined to reconstruct a three-dimensional reciprocal-space volume\cite{arnold_mantiddata_2014,michels-clark_expanding_2016}.

The crystal orientation was subsequently adjusted to access the [0,0,l] direction in reciprocal space in order to probe the temperature dependence of the observed [0,0,0.5]-type reflections. The sample was then warmed at a rate of 0.2 K min$^{-1}$ to temperatures above 18 K. A second rotation scan was performed at 15 K, where magnetic order was absent.

The integrated intensities were corrected for strong absorption effects using a model based on the sample geometry\cite{arnold_mantiddata_2014}. The average crystal structure was refined against nuclear Bragg reflections collected at 5 K using Jana2020\cite{petricek_jana2020_2023}.

Magnetic structure analysis was performed with the Bilbao Crystallographic Server\cite{perez-mato_symmetry-based_2015}. Using the propagation vector and the Eu site (Wyckoff position $(0,0,0)$) in the $P6_3/mmc$ space group, candidate magnetic space groups were systematically generated down to the lowest symmetry. Antiferromagnetic (AFM) models with finite moments on all sites were subsequently tested within Jana2020. Notably, configurations with moments constrained strictly along the $c$-axis cannot account for the observed $00l$-type magnetic reflections.

\section{First-principles calculations}

We present details of the first-principles calculations we have carried out. VASP calculations \cite{kresse1996efficiency,kresse1996efficient,kresse1999ultrasoft} were performed on primitive unit cells of the P6$_3$/mmc space group. The lattice constants used are $a=4.50 \AA$ and $c= 8.09\AA$ which also correspond to the relaxed lattice constants. In VASP calculations, we primarily used the \textsc{potpaw.64} set. Structural relaxation was carried out with Eu $f$ electrons frozed in core, and the valence of Eu was fixed to +2. We used the hardest potentials available for all atoms, in structural calculations. A cutoff of $E_c=450 \textrm{eV}$ was used. All forces were converged to be less than $1 \textrm{meV}/\AA$. After structural relaxation, SOC was added to determine the final total energy. For all calculations with $f$ electrons, a Hubbard U was on the f orbitals, $U=6.0 \textrm{eV}$. The k-point grid used throughout was $11 \times 11\times 6$. A $3\times 3 \times 3$ was used in all-electron calculations with SOC (see main text).

In determining the energy ladder for the phases in Table 1 (main text), we observed differences between the .54 and .64 pseudopotentials. For .54, the lowest energy phase was in fact FM-x. We attribute this to the different choice of the pseudoized orbitals, $4f^{6.5} 5s^2 5p^6 5d^{0.5} 6s^2$ in .64 and $ 4f^7 5s^2 5p^6 6s^2$ in .54.  To extract the surface spectrum, we Wannierized the AFM bands using Wannier90 \cite{Mostofi2014}, projecting on Eu-$d$, Ag-$d$, As-$p$ orbitals. The surface states of AFM EuAgAs were computed using WannierTools \cite{WU2017}. 3D Fermi surface contours were obtained using IFermi \cite{ganose2021ifermi}.

\begin{acknowledgments}
We thank Yujia Teng, Karin M Rabe, Andrea Urru and David Vanderbilt for helpful discussions, and particularly for Refs.~\cite{Urru2025,Teng2025}.
N.J.G acknowledges the support of the NSF CAREER award DMR-2143903. I.I.M. acknowledges support from the National Science Foundation under Award No. DMR-2403804. DK is supported by the Abrahams Postdoctoral Fellowship of the Center for Materials Theory, Rutgers University. This research used resources at the Spallation Neutron Source, a DOE Office of Science User Facility operated by the Oak Ridge National Laboratory. The beam time was allocated to CORELLI on proposal number IPTS-36263.1.

\end{acknowledgments}

\textbf{Author Contributions:} N.J.G. conceived and coordinated the project. M.E.G. grew the crystals. M.E.G. characterized the samples. M.E.G. performed the magnetic and magnetotransport measurements. R.R. and A.N. contributed to the magnetic and transport measurements. Z.M. and H.C. carried out neutron diffraction experiments. D.K. and I.I.M. carried out first-principles calculations and theoretical modelling. M.E.G, D.K. and N.J.G wrote the manuscript with contributions from I.I.M. All authors contributed to the discussion of the results. 

\textbf{Competing Interests:} The authors declare that they have no competing interests. 

\textbf{Data and Materials Availability:} All data needed to evaluate the conclusions in the paper are present in the paper and/or the Supplementary Materials. Additional data related to this paper may be requested from the authors.


\bibliography{main.bbl}

\renewcommand\thesection{\arabic{section}}
\renewcommand{\theequation}{S\arabic{equation}}
\setcounter{equation}{0}
\setcounter{figure}{0}
\setcounter{table}{0}
\setcounter{page}{1}
\makeatletter
\renewcommand\thesection{S\arabic{section}}
\renewcommand{\theequation}{S\arabic{equation}}
\renewcommand{\thetable}{S\arabic{table}}
\renewcommand\thefigure{S\arabic{figure}}
\renewcommand{\theHtable}{S\thetable}
\renewcommand{\theHfigure}{S\thefigure}

\clearpage
\onecolumngrid
\begin{center}
    \Large \textbf{Supplemental information for:\\ Near-degenerate competing magnetic orders in EuAgAs: a tunable route to altermagnetism} \\
    \vspace{10pt}
\end{center}

\begin{table}[h]
\caption{Crystallographic data, atomic coordinates and equivalent displacement parameters for $\text{EuAgAs}$.}
\label{ST1}
\renewcommand{\arraystretch}{1}
\small
\centering
\begin{tabular}{l l}
\hline
Crystal system & Hexagonal \\
Space group & $P6_3/mmc$ \\
Temperature (K) & $250.00$ \\
Wavelength (\AA) & $0.71073$ \\
Z formula units & 2 \\
\textit{a} (\AA) & $4.5052(2)$ \\
\textit{b} (\AA) & $4.5052(2)$ \\
\textit{c} (\AA) & $8.0929(5)$ \\
Volume (\AA$^3$) & $164.2601(9)$ \\
Goodness-of-fit on $F^2$ & $1.208$ \\
Final R indices [I $\geq$ 4$\sigma$(I)] & $\text{R}_1 = 0.0130$ \\
All data & $\text{R}_1 = 0.0137$, $\text{wR}_2 = 0.0405$ \\
Largest diff. peak \& hole (e \AA$^{-3}$) & $0.606$ and $-1.088$ \\
\hline
\end{tabular}

\vspace{0.4cm}

\centering
\begin{tabular}{c c c c c c}
\hline
Atom & Wyck. & \textit{x} & \textit{y} & \textit{z} & $\text{U}_{\text{eq}}$ (\AA$^2$) \\
\hline
$\text{Eu1}$ & 2a & $0.000000$ & $1.000000$ & $0.500000$ & $0.00972$ \\
$\text{Ag1}$ & 2c & $0.333333$ & $0.666667$ & $0.250000$ & $0.00865$ \\
$\text{As1}$ & 2d & $0.666667$ & $0.333333$ & $0.250000$ & $0.00559$ \\
\hline
\end{tabular}
\end{table}

\begin{figure*}
    \centering
    \includegraphics[width=0.9\linewidth]{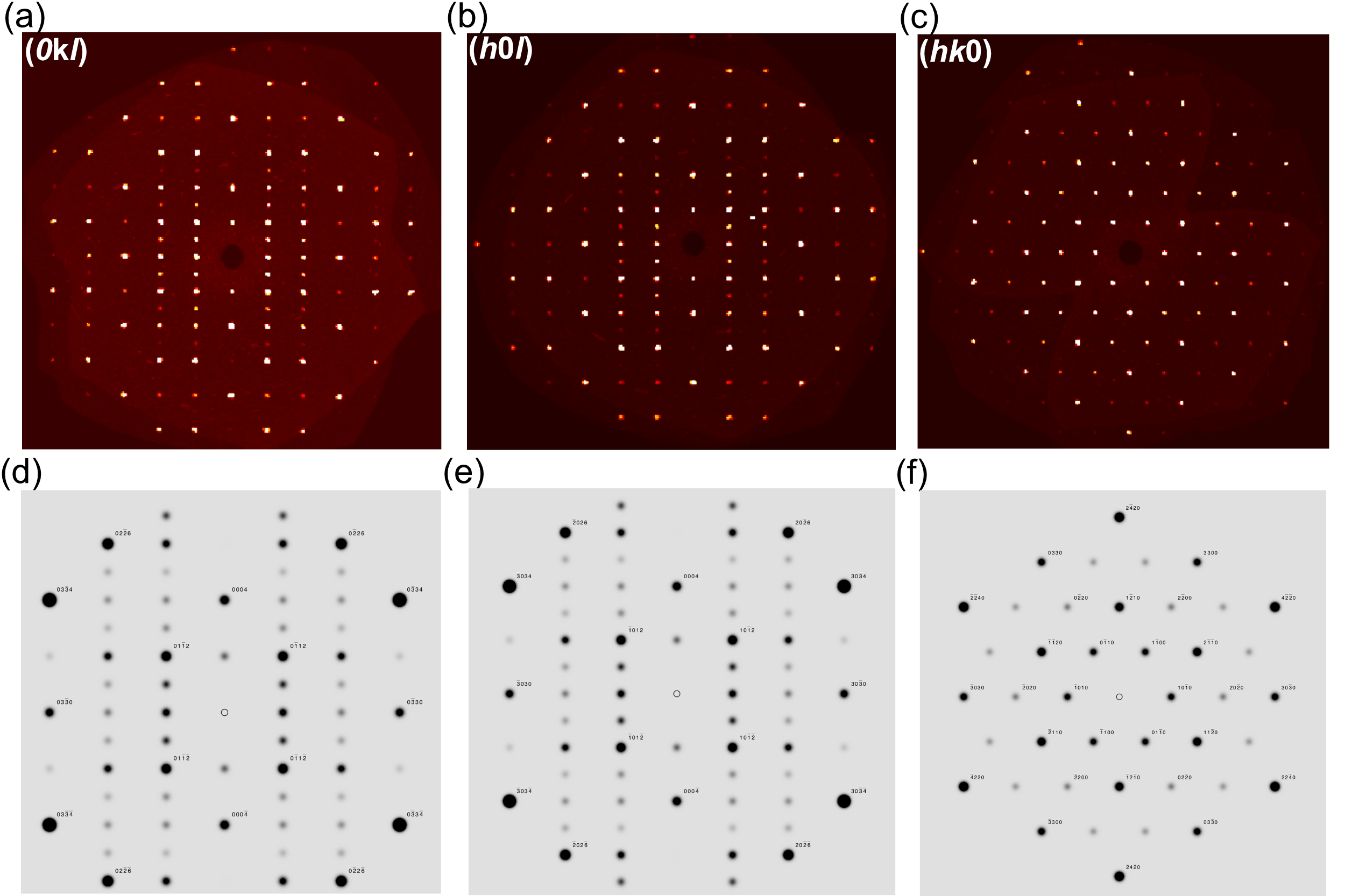}
    \caption{\textbf{Experimental and simulated reciprocal-space planes from single-crystal X-ray diffraction.}
    (a-c) Experimental X-ray single crystal diffraction precession images in the $(0kl)$, $(h0l)$, and $(hk0)$ planes, respectively. (d--f) Simulated diffraction patterns corresponding to panels (a--c).}
    \label{FigS1}
\end{figure*}

\begin{figure*}
    \centering
    \includegraphics[width=1\linewidth]{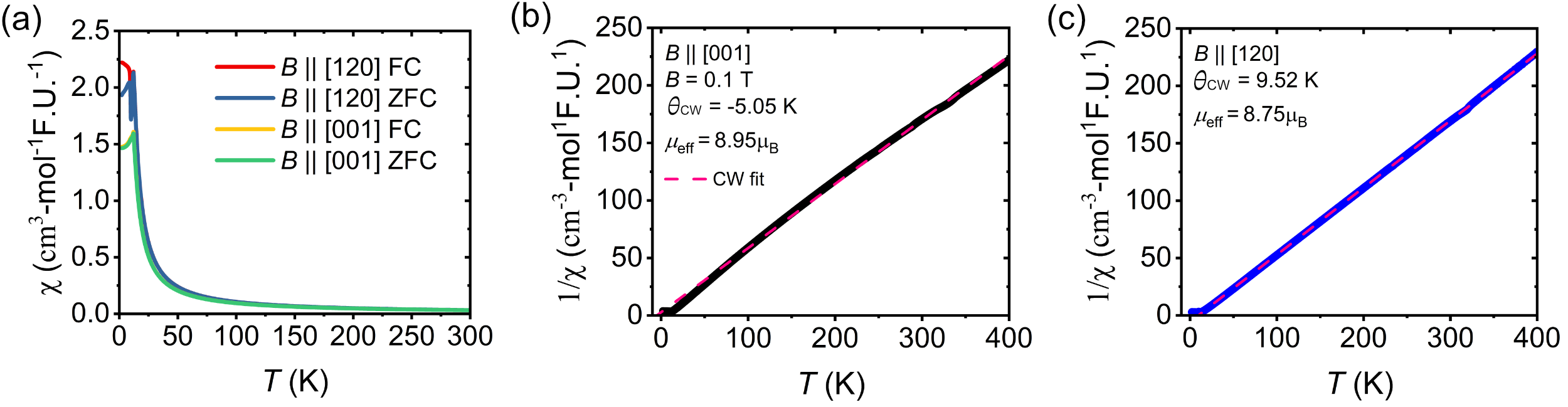}
    \caption{ \textbf{Temperature-dependent magnetic susceptibility and Curie-Weiss analysis.}
    (a) Zero-field-cooled (ZFC) and field-cooled (FC) magnetic susceptibility $\chi(T)$ for $B \parallel [120]$ and $B \parallel [001]$. (b) Inverse susceptibility $1/\chi(T)$ for $B \parallel [001]$, together with a linear Curie--Weiss fit to the high-temperature paramagnetic regime, yielding $C = 10.02~\mathrm{cm^3\,K\,mol^{-1}}$, $\mu_{\mathrm{eff}} = 8.95~\mu_B$ per formula unit, and $\theta_{\text{CW}}=-5.05$ K. (c) Inverse susceptibility $1/\chi(T)$ for $B \parallel [120]$, together with a linear Curie--Weiss fit, yielding $C = 9.56~\mathrm{cm^3\,K\,mol^{-1}}$,  $\mu_{\mathrm{eff}} = 8.75~\mu_B$ per formula unit, and $\theta_{\text{CW}}=9.52$ K.}
    \label{FigS2}
\end{figure*}

\begin{figure*}
    \centering
    \includegraphics[width=1\linewidth]{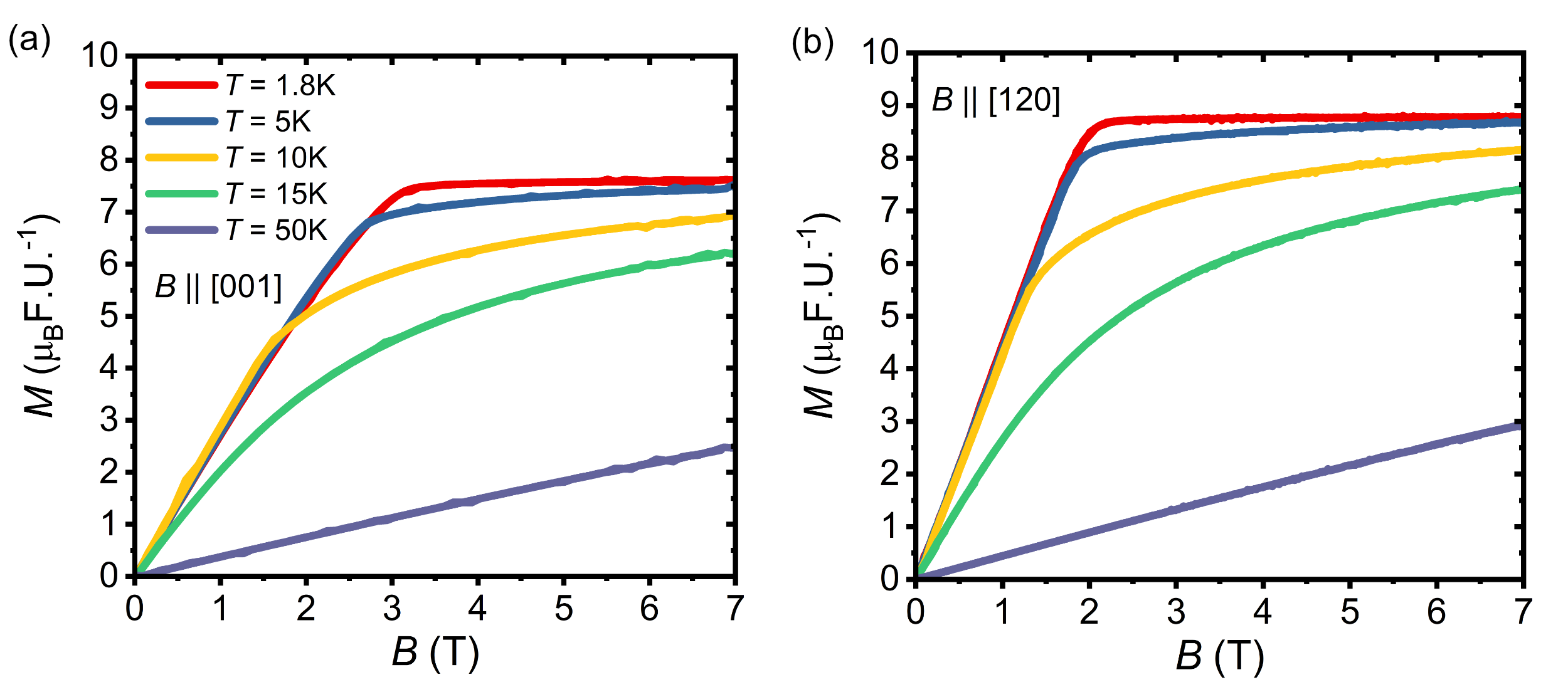}
    \caption{\textbf{Field-dependent magnetization.}
     $M(B)$ measured for (a) $B \parallel [001]$ and (b) $B \parallel [120]$ at $T=1.8-50$ K.}
    \label{FigS3}
\end{figure*}

\begin{figure*}
    \centering
    \includegraphics[width=1\linewidth]{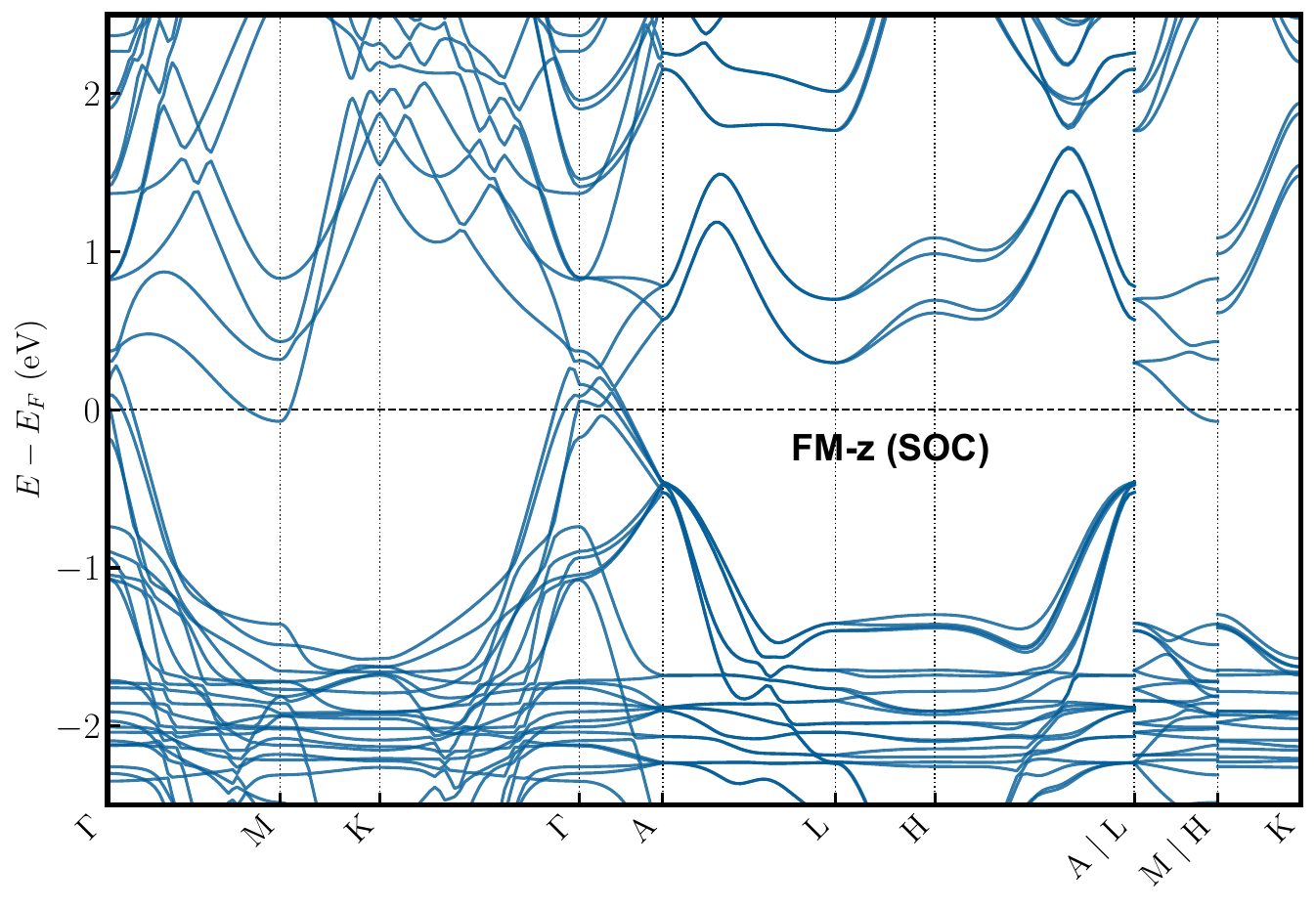}
    \caption{Band structure from the FM-z phase of EuAgAs with SOC. All qualitative features of the band structure without SOC are preserved, including well-defined pockets near $M$ and $\Gamma$. Small gaps appear along the $\Gamma-A$ line, but Weyl points created from splitting the Dirac point due to magnetic order remain gapless.}
    \label{FigS4}
\end{figure*}

\begin{figure*}
    \centering
    \includegraphics[width=1\linewidth]{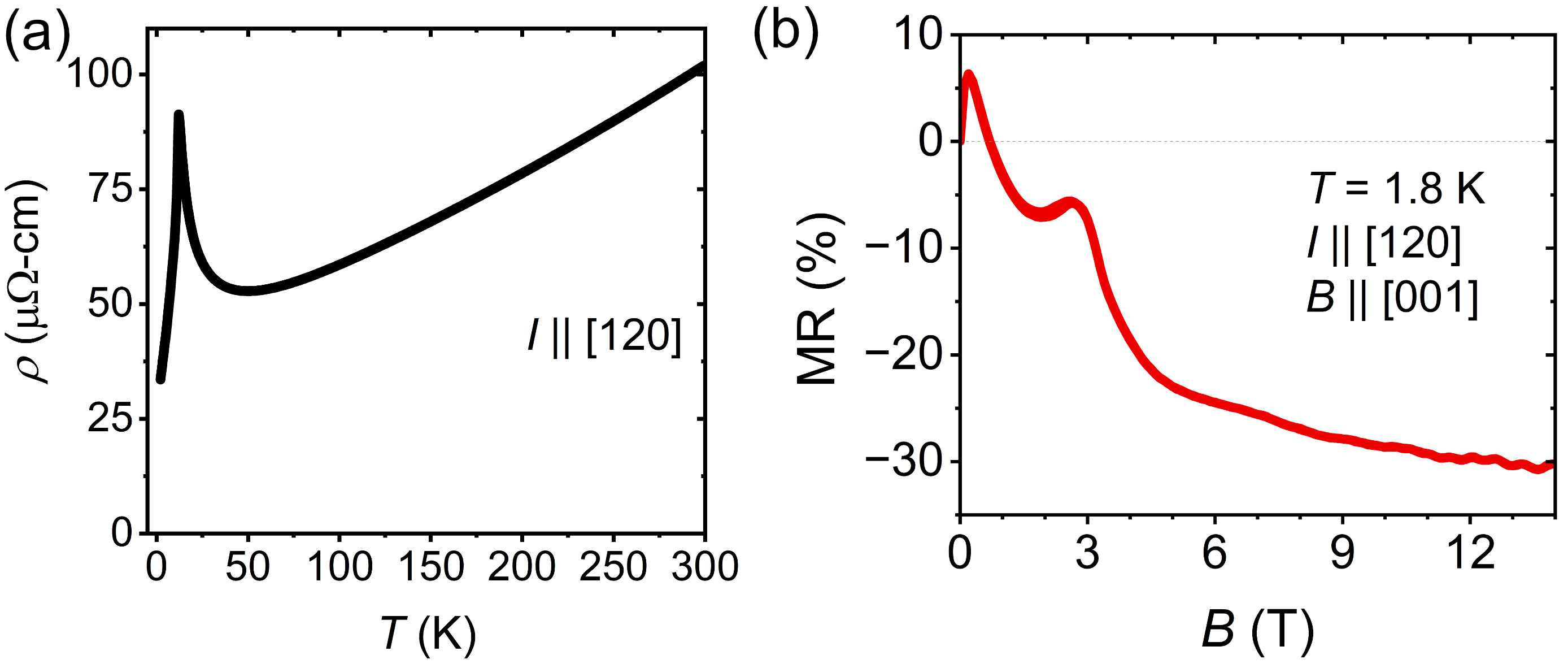}
    \caption{\textbf{Electrical transport.}
    (a) Temperature dependent longitudinal resistivity of EuAgAs, measured along the $[120]$ direction. (b) Magnetoresistance of EuAgAs measured at $T=1.8$ K, for $I$ $||$ [120] and $B$ $||$ [001], highlighting the appearance of Shubnikov-de Haas oscillations above 8 T.}
    \label{FigS5}
\end{figure*}

\end{document}